\documentclass[twocolumn,prl, aps,superscriptaddress,showpacs]{revtex4-1}

\usepackage{mathptmx}
\usepackage{subfigure}
\usepackage{dcolumn}
\usepackage{amsmath,amssymb}
\usepackage{bm}
\usepackage{color}
\usepackage{latexsym}
\usepackage{epstopdf}
\usepackage{color}
\usepackage[english]{babel}
\usepackage{latexsym}

\usepackage{soul}

\usepackage{psfrag,graphicx} 
\usepackage{epsf} 
\usepackage{subfigure} 
\usepackage{amsmath} 
\usepackage{amssymb} 
\usepackage{amsfonts}
\usepackage{bm}
\usepackage{natbib}
\usepackage{epstopdf}
\usepackage{appendix}

\definecolor{mygrey}{gray}{0.35}
\definecolor{myblue}{rgb}{0.2,0.2,0.8}
\definecolor{myzard}{cmyk}{0,0,0.05,0}
\definecolor{mywhite}{rgb}{1,1,1}
\definecolor{myred}{rgb}{1,0.,0.3}

\usepackage[colorlinks=true,citecolor=myblue]{hyperref}

\def\be{\begin{equation}}
\def\ee{\end{equation}}
\def\ba{\begin{align}}
\def\enda{\end{align}}
\def\bi{\begin{itemize}}
\def\ei{\end{itemize}}

 \def\ee{\mathord{\rm e}}

 \def\ee{\mathord{\rm e}}

\renewcommand{\ee}{{\rm e}}

\def\beq{\begin{equation}}
\def\beq{\begin{equation}}
\def\eeq{\end{equation}}

 \newcommand{\ket}[1]{|#1\rangle}
 \newcommand{\bra}[1]{\langle #1|}

\binoppenalty=\maxdimen
\relpenalty=\maxdimen
\begin{document}

\title[Short Title]{Increasing sensing resolution with error correction}

\author{G. Arrad}
\author{Y. Vinkler}
\author{D. Aharonov}
\author{A. Retzker}
\affiliation{Racah Institute of Physics, The Hebrew University of Jerusalem, Jerusalem 91904, Givat Ram, Israel}
\date{\today}

\date{\today}

\begin{abstract}
{The signal to noise ratio of quantum sensing protocols scales with the square root of the coherence time. Thus, increasing this time is a key goal in the field.  Dynamical decoupling has proven to be efficient in prolonging the coherence times for the benefit of quantum sensing. However, dynamical decoupling can only push the sensitivity up to a certain limit. In this work we present a new approach to increasing  the coherence time further through error correction which can improve the efficiency of quantum sensing beyond the fundamental limits of current state of the art methods.}
\end{abstract}

\maketitle

Quantum sensing and metrology\cite{Lloyd} are key goals of quantum technologies. Impressive  achievements have been made in both in recent years.
The frequency uncertainty of atomic clocks has decreased dramatically\cite{clock1,clock2}, the signal to noise ratio of magnetic field measurements has considerably increased\cite{Balasubramanian,Hong,polzik2,kotler,maze} and the contrast of spin imaging has improved \cite{Staudacher,london}.
Since the sensitivity of quantum sensing scales as $\frac{1}{\sqrt{T_2}},$ where $T_2$ is the coherence time, a great deal of effort has been devoted to designing and realizing protocols to increase this time while also maintaining the sensing signal.  However, the state of the art protocols, which are based on dynamical decoupling(DD),  can only tackle low frequency noise since the DD control has to be faster than the correlation time of the noise.  Thus DD  can only increase the coherence time up to a certain limit.  This limit can be overcome by the use of error correction(EC), which need not be faster than the noise correlation time, only its effect, which is much slower.
Therefore, the utilization of EC for quantum sensing objectives could increase the signal to noise ratio substantially, and thus enhance the sensitivity of field measurement and the contrast of imaging.

\begin{figure}
   \centering
   \includegraphics[width=1\columnwidth]{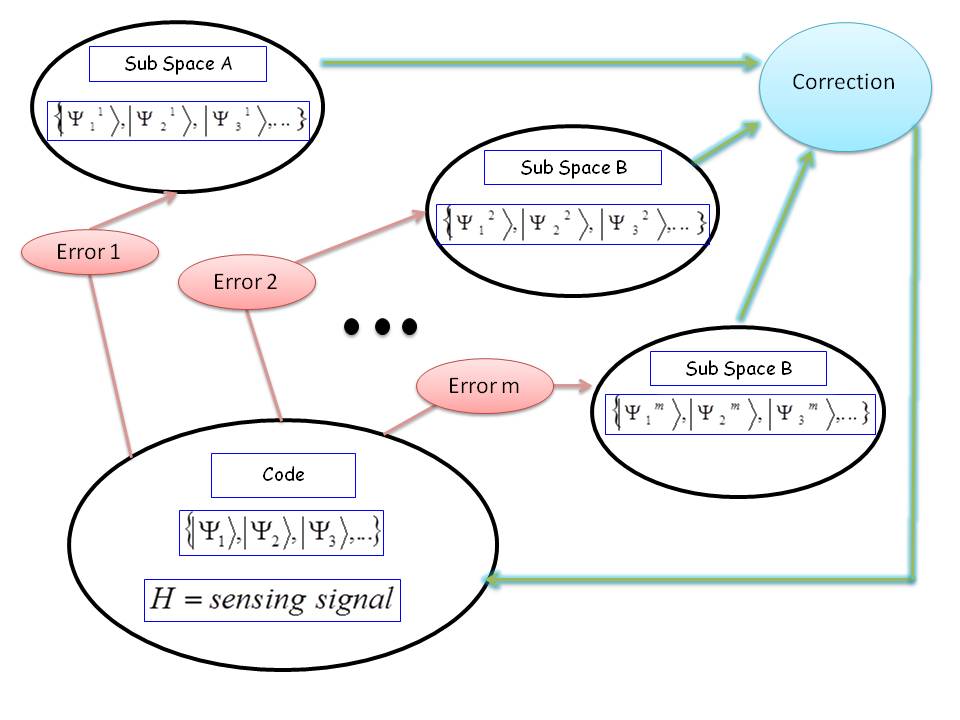}
	\caption {The basic mechanism of the combination of EC with sensing. Errors map the code space to orthogonal spaces, and is followed by EC sequence, composed of a detection and then a correction, that brings the state back to the code. }
  \label{scheme}
  \end{figure}

EC tackles high frequency noise by using redundant qubits.  Following Shor's work\cite{shor}, various protocols have been proposed, including Stean's code\cite{steane} and various fault tolerant methods\cite{daniel}. Recently, several EC protocols have been realized \cite{blat1,wrachtrup}.  Here we introduce the concept of using EC for quantum sensing and  present a set of protocols that achieve this goal. The basic idea is shown in fig. \ref{scheme}. An EC scheme is composed of a code sub-space  $\{\vert\psi_1, \psi_2,...\psi_N\rangle\}$, in which all the pertinent information is found; i.e., the sensing signal should work inside the code (e.g. $H_{s}=g\vert\psi_i\rangle \langle\psi_j \vert+h.c.$,  where throughout this article $g$ is the signal).
The code is susceptible to errors, which map it to orthogonal subspaces. Correction of the errors is done by means of projective measurements of the various subspaces, and applying a correction sequence.

There are two main differences between EC for quantum computing and EC for sensing. While a logic operation can be realized by fast arbitrary pulses  that rotate between the code states, the sensing signal for most scenarios is weak, continuous, and very specific. This difference is the basis for the main complication in the schemes  presented here. However, the sensing mechanism can also potentially benefit from the use of fully protected qubits, which are neither sensitive to noise nor to the measured fields. Thus, we will make use of one or more qubits which are 
assumed to be "good", in contrary to these that are sensitive to the signal. These 'good' qubits could be  produced by clock states or robust nuclear spins. There is no analog for this in the case of quantum computing since in that case we use, by design, the most robust qubits available. These two characteristics  thus define the main features of the sensing codes.

{\em Improving dynamical decoupling with error correction ---}
DD fails in two main scenarios. The first is when DD is not fast enough to overcome the correlation time of the noise and the second is when DD suffers from noise in the control. We address these issues below.   We start by presenting a few physically relevant sensing models in which the effects of errors can be ameliorated by an appropriate error-correction protocol. 

Classical Drive Noise ---  The first model we consider is comprised of a single Two Level System (TLS), composed of the basis states $|\!\uparrow\rangle$ and $|\!\downarrow\rangle$, which are separated by an energy gap $\omega_0$, and are driven by an external drive $\Omega$ at frequency $\omega_0$. The sensing signal $g$ is coupled to the TLS and the system is described by the Hamiltonian
\[
	H = \left[\frac{\omega_0}{2}+f(t)\right]\sigma^z+
						\left(\Omega+\delta\Omega\right)\sigma^x\cos(\omega_0 t) + 
						g\sigma^z\cos(\Omega t).
\]
Here $f(t)$ and $\delta\Omega$ represent the external and the control noise and $\sigma^i$ are the Pauli operators. This Hamiltonian represents the main magnetometry scenario which has been realized in NV centers\cite{maze,Balasubramanian}, ions\cite{kotler} and atoms\cite{polzik1}. By transferring first to the interaction picture with respect to $\omega_0\sigma^z/2$ and then to the interaction picture with respect to $\Omega\sigma^x$, assuming that $f(t)\ll\Omega \ll \omega_0$ and taking advantage of the rotating-wave-approximation, we are left with the Hamiltonian: $	H_{\rm I} = \frac{g}{2}\sigma^z+\delta\Omega\sigma^x.$

\begin{figure}
   \centering
   \includegraphics[width=1\columnwidth]{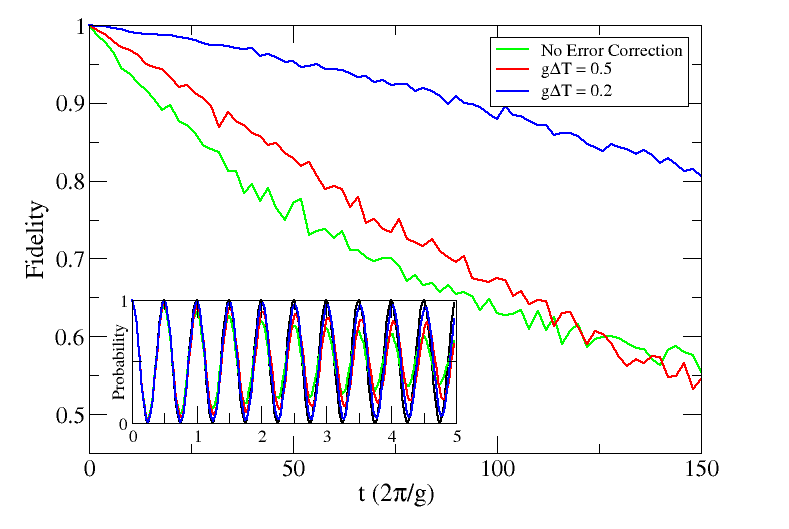}
	\caption {(Color online) The fidelity of a state as a function of time, for different time differences between EC operations. Here each point represents an average over $N=1024$ simulations, and the fidelity of the system is plotted for run times that are integer multiplications of $2\pi/g$. In each run either no EC procedure was applied (green), an EC procedure was applied each time interval of $g\Delta T = 0.5$ (red) or at time intervals of $g\Delta T = 0.2$ (blue). In each case the range of the randomly chosen noise was $(-g/2,g/2)$. Inset: The probability of measuring the initial state of the system, as a function of time. Here the black line represents the case with no noise, and the different lines show the case where no EC was made (green), or where an EC was made at intervals of $g\Delta T = 0.5$ (red) and $g\Delta T = 0.2$ (blue). Here we considered a stronger noise chosen randomly within the range of $(-2g,2g)$.}
  \label{sim1}
  \end{figure}

We designate the eigenstates of $\sigma_z$ as $\vert \downarrow \rangle $ and $\ket \uparrow $, and we  assume that we have a good qubit $\left\{ \ket 0, \ket 1 \right\}$, with a known energy gap $\nu,$ that can be controlled at will. The code is defined as:
\begin{equation}
	\{ \vert \downarrow 0 \rangle, \vert \uparrow 1 \rangle \},
	\label{code_classical}
\end{equation}
the signal $g$  inflicts a phase shift between the two states which may be detected and the error $\delta\Omega$ represents a bit-flip operation on the first qubit, taking the system into the orthogonal subspace
$\{ \vert \uparrow 0 \rangle, \vert \downarrow 1 \rangle \},$
and thus a projective measurement could correct the error. However, the noise is only correctable at a  short time scale with respect to $1/g$ since the sensing tends to rotate the noise, hence generating an effective general noise that cannot be corrected by the code. 
We have checked that numerically (fig. \ref{sim1}) and have seen that indeed for a high EC repetition rate, the error is correctable.

Noise in all directions --- 
In case of general noise , the system  Hamiltonian is described by:
\[
H=f_{z}\sigma_{z}+f_{x}\sigma_{x}+f_{y}\sigma_{y}+\Omega\sigma_{x}+g\sigma_{z}\cos\Omega t,
\]
and in the interaction picture with respect to the drive, in the limit when $\Omega$ is much faster than the noise, it is approximated by:
\[
H=f_{x}(t)\sigma_{x}+\frac{g}{2}\sigma_{z},
\]
and this noise can be dealt with by the previous code. 

This method can be incorporated in  a pulsed DD scheme as well.
Suppose we have a system with the following Hamiltonian
\[
H_{sense}=g\sigma_{z}\cos\omega_{0}t +f_{z}(t)\sigma_{z}+f_{x}(t)\sigma_{x},
\]
 where we estimate $g_{0}\simeq g$ ,i.e.  $g=g_{0}+\Delta g,$  and we aim to evaluate the correction $\Delta g$. The $f_i(t)$ represents the noise, where $f_x(t)$ is assumed to be fast and thus cannot be dealt  with by DD.
In order to correct this, we can work with the previous code:$\left\{ \vert\uparrow1\rangle,\vert\downarrow0\rangle\right\}.$  The $f_{x}(t)\sigma_{x}$ term can be dealt with by EC as was the case in the previous scheme, and the  $\sigma_{z}$ term can be dealt with  by pulsed DD the following way.

The sensing signal will induce a phase shift between the two states. Assuming that the measurement is repeated every time $\tau$ which is followed by a $\pi$ pulse switching the population of the two code states, the following state is realized: $
\vert\uparrow1\rangle+e^{-i\int_0^t g cos\left(\omega_{0}t\right) dt}\vert\downarrow0\rangle\to e^{-i\frac{g}{\omega_{0}}sin\left(\omega_{0}t\right)}\vert\uparrow1\rangle+\vert\downarrow0\rangle.$
The $\ket{\downarrow0}$ continue acquiring the phase $\phi=\int_\tau^{2\tau} g cos\left(\omega_{0}t\right) dt$ until the next pulse.
Suppose we choose $\tau=\frac{\pi}{\omega_0}$ , and repeat the same procedure $n=t/\tau$ times. Note that: $\sum_{k=0}^n\int_{2n\tau}^{(2n+1)\tau} g cos\left(\omega_{0}t\right) dt=n\frac{g}{\omega_{0}}=t\frac{g}{\pi}$ we gain the phase in a linear process:
\[
\ket{\psi_t}=e^{-i\frac{g}{\pi}t}\vert\uparrow1\rangle+e^{i\frac{g}{\pi}t}\vert\downarrow0\rangle,
\]
This procedure illustrates the pulsed version of the combination of DD and error
correction. In case that a noise in the $y$ direction exists as well, part of it will be corrected by error correction and part by the DD.

Measuring interaction ---  The next model we present is composed of two identical two-level systems with energy gaps $\omega_0,$ which are coupled by the sensing signal $g,$ as described by the Hamiltonian
\begin{equation}
	H = \sum_j\left[\frac{\omega_0}{2}+f_j(t)\right]\sigma_j^z
						+\frac{g}{2}\sigma^x_1\sigma^x_2,
\end{equation}
where $f_j(t)$ represents noise in the same direction as the energy gap, and we have arbitrarily taken the signal to be $g/2$ so that it coincides with the previous model. 

The Hamiltonian divides the physical space into two disconnected subspaces, the first spanned by the states $|\downarrow_1,\uparrow_2\rangle$ and $\ket{\uparrow_1,\downarrow_2}$ and the second one spanned by the states $\ket {\downarrow_1,\downarrow_2}$ and $\vert\uparrow_1,\uparrow_2\rangle$. Going into the interaction picture with respect to $\omega_0\left(\sigma^z_1+\sigma^z_2\right)/2$ the dynamics of the former subspace are described by the interaction Hamiltonian
\begin{equation}
	H_{\rm I} = \sum_j f_j(t)\sigma_j^z + \frac{g}{2}\left(
					\sigma_1^+\sigma_2^-+{\rm h.c.}\right).
\end{equation}
Working in that subspace, we note that the states $\left( \vert \downarrow_1,\uparrow_2\rangle \pm \vert\uparrow_1,\downarrow_2\rangle \right)/\sqrt{2}$ are eigenstates of the signal part of the Hamiltonian $(\sigma_1^+\sigma_2^-+{\rm h.c.})$ with eigenvalues $(-1)$ and $(+1)$, respectively. We again use the existence of a 'good' qubit $\left\{ \ket 0,\ket 1\right\}$ with energy gap $\nu$ in order to define the code states in an identical manner to the ones defined in Eq.~(\ref{code_classical}):
\begin{equation}
	\left\{\frac{\vert \downarrow_1,\uparrow_2,0\rangle 
			-\vert\uparrow_1,\downarrow_2,0\rangle}{\sqrt{2}},
			\frac{\vert \downarrow_1,\uparrow_2,1\rangle 
			+\vert\uparrow_1,\downarrow_2,1\rangle}{\sqrt{2}}
	\right\}.
\end{equation}
The error ($E_j=\sigma^z_j$) maps the code to an orthogonal subspace, allowing for a correction by projective measurement.

{\em Prolonging $T_1$ ---}  In order to deal with decay errors, i.e. $T_1$ errors, we need a more elaborate EC scheme. One can use traditional codes, which assume that an error can occur in any of the qubits. However, in order to utilize these codes for sensing, extremely sophisticated protocols are required. We propose much simpler codes that use 'good' qubits. By constructing these codes we develop the main ideas  which could also be used as building blocks for combining sensing with traditional codes.
Note that dealing with $T_1$ noise is not the same as dealing with general errors. (The generality of $T_1$ errors is discussed in \cite{Sup} )

Sensing becomes more challenging for $T_1$ noise since a specially tailored protocol should be used to distinguish the signal from the noise. Specifically, the code states must differ from each other by the state of at least two of the qubits. The most obvious way to do this is with a Raman transition between two codes states via a state which is outside the code, a utility state. As the utility state is not inside the code, errors on it will not be correctable. Moreover, some errors can directly connect the code and the utility state. (see \cite{Sup})

The following protocols manage to overcome the noise and enlarge coherence times because two factors come into play: 1) The small population of the utility state. 2) The special characteristics of the dissipation noise.  The main problem with this protocol is explained below.

\begin{figure}
   \centering
 \includegraphics[height=3.5cm,width=8cm]{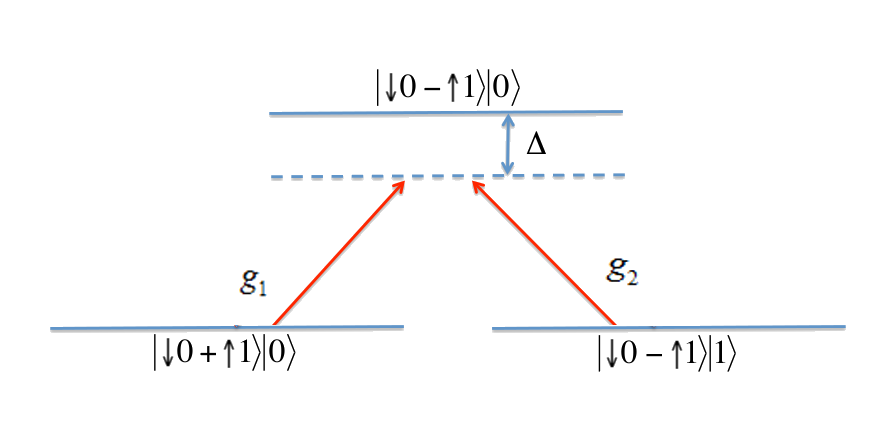}
	\caption {The level structure of the basic model to overcome general noise. $g_1$ is the sensing signal while $g_2$ is the external driving.   }
  \label{Raman1}
  \end{figure}

In this error model the first qubit has a limited $T_{1}$ time; in other words, it is susceptible to dissipation, whereas the other two are 'good' qubits. The proposed code is:
\[
   \vert A\rangle=\vert\downarrow 0+\uparrow1\rangle\vert0\rangle \hspace{0.25cm};\hspace{0.25cm}   \vert C\rangle=\vert \downarrow 0-\uparrow 1\rangle\vert1\rangle
\]
which is fully correctable; i.e.,  both bit flips and phase flips can be corrected. 
The sensing signal, however,  is a phase flip itself and thus a scheme which distinguishes between the noise and the signal is needed.
Let's look at the following procedure. By opening a gap between these
two states and the  utility state $\vert B\rangle=\vert \downarrow 0 - \uparrow 1\rangle\vert 0 \rangle$
, a Raman transition mediating the sensing signal between the code states,  can be designed.  The level structure is shown in fig. \ref{Raman1}.
The Hamiltonian is: $ H_s=g_1 \vert B\rangle\langle A \vert+ g_2 \vert B\rangle\langle C \vert+h.c.+\Delta\vert B\rangle \langle B \vert $.
In this case the utility state will always be occupied by  the small amplitude $\epsilon=\frac{\Omega}{\delta}$.
Suppose we arrive, at time t, at that state:
\begin{equation}
\left(\alpha\vert00-11\rangle\vert1\rangle+\beta\vert00+11\rangle\vert0\rangle+e^{it\Delta}\epsilon\vert00-11\rangle\vert0\rangle\right)\vert n=0\rangle,
\end{equation}
The emission of a photon will result in:
$ \left[\alpha\vert00\rangle\vert1\rangle+(\beta+e^{it\Delta}\epsilon)\vert00\rangle\vert0\rangle\right]\vert n=0\rangle+[-\alpha\vert01\rangle\vert1\rangle+...$
 $...+(\beta-e^{it\Delta}\epsilon]\vert01\rangle\vert0\rangle)\vert n=1\rangle,$
Measuring $S_z^1 S_z^2$ , i.e. separating according to different $n$ populations, and correcting, we get:
 $\alpha\vert00+11\rangle\vert0\rangle+(\beta\pm e^{it\Delta}\epsilon)\vert00-11\rangle\vert1\rangle$
 \cite{Sup} .

This shows that (for $\Delta T_1 \gg 1$) $\beta$ undergoes a random walk and thus
correction to the signal would only result in second order terms, giving us a longer 
coherence time. The uncertainty, however, stays the same as the original uncertainty $\delta g_0 =\frac{1}{\sqrt{T_{1}}}$ (up to a small numerical factor) since:
$\delta g_{Raman}=\frac{\delta}{\Omega}\frac{1}{\sqrt{T_{2}}}=\frac{1}{\sqrt{T_{1}}}$, where the first equation is due to the Raman transition's slow effective rotation frequency of $\omega_{eff}=\frac{g\Omega}{\delta}$, and in the second we substituted: $\sqrt{T_{2}}=\frac{\delta}{\Omega}\sqrt{T_{1}} $ which we derived from the condition that the variance of the noise random-walk reaches unity  $1=\epsilon\sqrt{N}=\frac{\Omega}{\delta}\sqrt{\frac{T_{2}}{T_{1}}}$.

{\em Working in the strong noise regime --- }
A possible application for the above scheme can be found in systems with a $T_1\cdot g \ll 1$.
The precision of a frequency measurement for a system with cosine signal $P_{1}=\cos(g\cdot t)$ and $T_{1}$ decay time, is shown here
\cite{martin} to depend on time as:
\begin{equation}
\delta g=\frac{1}{2}\frac{\sqrt{e^{t/T_{1}}-\cos^{2}\left(2gt\right)}}{\sqrt{nTt}\sqrt{\sin^{2}\left(2gt\right)}},
	\label{dg_of_t}
\end{equation}
where T is the total experiment time and $n$ the number of simultaneously running systems. 
The optimal time is achieved when $\sin^{2}(2gt)\simeq1$; i.e., near the middle of the cosine's period, which will give us essentially $\delta g\propto\frac{1}{\sqrt{T_{1} T n}}$ \cite{martin}. 
But, in the strong noise regime, $T_{1}\cdot g\ll1$, so we are restricted to the top of the cosine, where $\sin^{2}(2gt)\simeq (2gt)^{2}$. Substituting into eq.[\ref{dg_of_t}] differentiating and solving for best timing, we arive at $t_{max}=T_{1}$ giving us: \hspace{0.1cm}$\delta g_{strong}\propto g\cdot T_{1}^{-3/2}$ \hspace{0.1cm}, 
which is worse by a factor of $T_1 g$ \cite{Sup} .

One way to prevent this is by making a high detuning. In such interferometry our signal becomes $g_{eff}=g+\Delta$ ( where $\Delta$ is the detuning); thus we can get to the middle of the cosine even for signal g, and enhance the sensitivity. 
Still, this is not always simple to achieve, for example when measuring the strength of a weak drive, and in those cases we can use the above-mentioned scheme for EC. This EC will give us the sensitivity eq.[\ref{dg_of_t}], where we change $g\to \omega_{eff}=\frac{g\Omega}{\delta} $ and $T_1 \to T_2$ and now, using $T_{2}=\frac{\delta^{2}}{\Omega^{2}} T_{1}$, we get $\left(2\omega_{eff}T_{2}\right)= (2gT_{1}\frac{\delta}{\Omega}) \gg 1$. We can thus choose $\sin^{2}(2\omega_{eff}t)\simeq1$, which enables for the above-noted accuracy of $\delta g_{EC}\propto\frac{1}{\sqrt{T_{1}}}$.
Checking the relative accuracy we get:
\[
 \frac{\delta g_{EC}}{\delta g_{strong}}\propto T_{1}g \ll 1
\]

{\em Spin spin interactions ---  }
In this section we devise a scheme for measurement of the interaction strength between a dissipative TLS and a stable one. The dipole-dipole interaction will induce flip flops between the two.
Here we use the code: $\left( \vert \uparrow 0 0 \rangle + \vert \downarrow 11 \rangle  \right)/\sqrt 2$, $\left( \vert \downarrow 1 0 \rangle + \vert \uparrow 01 \rangle  \right)/\sqrt 2,$  where the first qubit is the sensing qubit and the other two are good qubits. This is a fully correctable code, as the error maps the code to orthogonal subspaces. Moreover, the flip flop interaction  $H=g\left( \sigma_+^1 \sigma _-^2 + h.c  \right) $  couples the two code states directly and thus the sensing protocol does not use a utility state which lies outside the code.
However, note that in order to use this we need to be able to apply non local interactions in the correction sequences  \cite{Sup}. This code could be useful in measuring spin - spin interactions between ions or measuring distances between an NV center and a nucleus.  

{\em   Measuring the sideband interaction term --- } 
The sideband interaction is the main building block for quantum information processing with trapped ions.
As the strength of the sideband interaction is proportional to the Rabi frequency, precise measurement of this term is analogous to measurement of laser, microwave or rf fields. 

Since the sideband interaction, $H=\eta \Omega \left( \sigma_- a^+  + h.c \right), $ where $\sigma_-$ is the raising spin operator and $a$ is the phonon distraction operator, creates flip flops, the previous code could be used when one of the good qubits is replaced by a phonon:
$\left( \vert \uparrow \rangle \vert n= 0 \rangle \vert 0 \rangle + \vert \downarrow \rangle \vert n = 1 \rangle \vert  1 \rangle  \right)/\sqrt 2$, $\left( \vert \downarrow \rangle \vert n = 1\rangle \vert  0 \rangle + \vert \uparrow \rangle \vert n =  0 \rangle \vert 1 \rangle  \right)/\sqrt 2.$
The precision of the measurement of this protocol is limited by the coherence time of the phonon.  \cite{Sup}.
This procedure could be used to measure Rabi frequencies and the Lamb-Dicke parameter.

\begin{figure}
  \centering
   \includegraphics[scale=0.2]{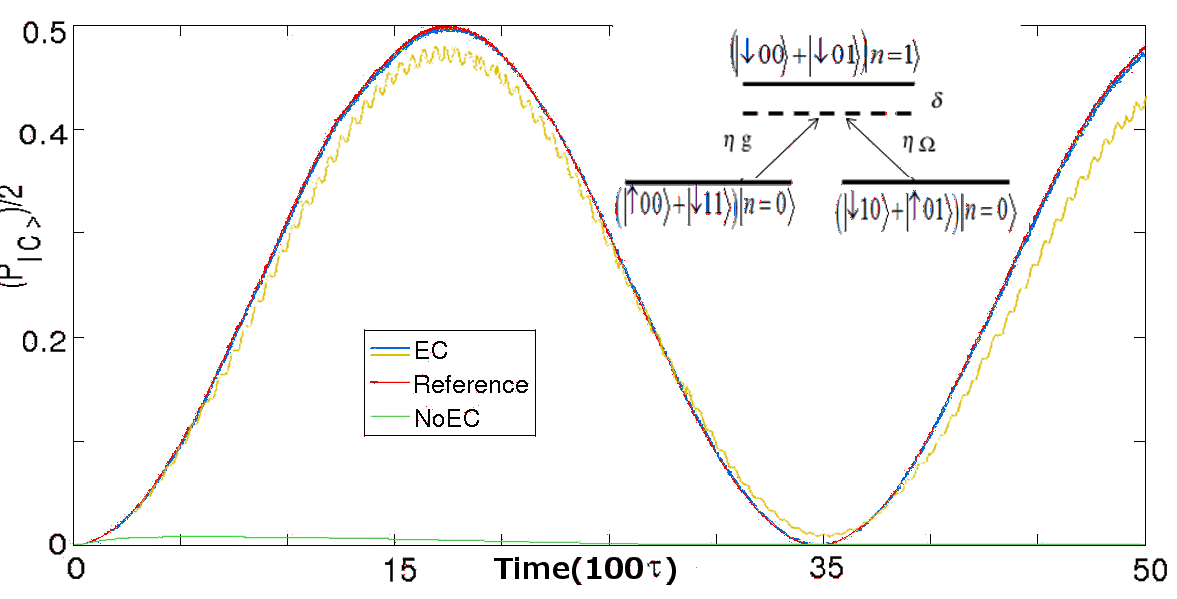}
	\caption {Red sideband interaction. The typical behavior of the red sideband scheme in the strong noise limit. The red line is the reference $\sin(\Omega t$). The blue line, which follows it almost exactly, is the simulated signal using EC with $T_1\tau\cdot=10^{-3}$ and $\frac{\Omega}{\delta}=10^{-2}$. The yellow line, which forms a rather crude sine signal, is the simulated signal using EC with $T_1\tau\cdot=10^{-2}$ and $\frac{\Omega}{\delta}=3\cdot10^{-2}$. The fast decaying green line  is the same simulation without EC and $\frac{\Omega}{\delta}=10^{-3}$. The inset represents the energy level diagram of the scheme.} 
  \label{MS}
  \end{figure}

{\em Molmer Sorensen coupling measurement ---  }
A natural question to explore is the utilization of MS gates for EC. As the MS coupling term scales as $\frac{\eta^2 g \Omega}{\delta}$ measurement, of this term provides additional information on top of the sideband; namely the detuning.  

The code states are $\vert A\rangle=\frac{\vert \uparrow00\rangle+\vert \downarrow11\rangle}{\sqrt{2}}\vert 0_{vib}\rangle$;
$ \vert C\rangle=\frac{\vert \downarrow10\rangle+\vert \uparrow01\rangle}{\sqrt{2}}\vert 0_{vib}\rangle$  and the ancillary  state is $\vert B\rangle=\frac{\vert \downarrow00\rangle+\vert \downarrow01\rangle}{\sqrt{2}}\vert 1_{vib}\rangle $.
The red side-band interaction is applied on both the first and the second ions, resulting in a Raman transition via the third state. The code is correctable; however, as the utility state is not correctable under phase flip the protocol is not perfect. We simulated this protocol to validate that a considerable gain exists, see fig.\ref{MS}  and \cite{Sup} .

{\em Utilizing super-radiance ---}
One way to measure the distance between two emitters that are closer than a wavelength apart is to measure the energy gap between the super radiant state and the irradiant state. Although the irradiant state has a longer lifetime and thus can be measured with higher accuracy, the supperradiant state has short lifetime and thus low accuracy.  The following procedure shows that by using dissipation we can still measure the energy gap with high accuracy.

The code is composed of the following two states: $ \vert A\rangle =\vert 111\rangle ,\vert B\rangle =\frac{\vert 010\rangle +\vert 100\rangle }{\sqrt{2}},$ with the Hamiltonian $H=g\left(\sigma_{z}^{1}+\sigma_{z}^{2}\right)+g \left( \sigma^1_+ \sigma^2_- +h.c \right)$ yielding an energy gap of $\Delta E=g+\omega_G$ which can be measured by Ramsey interferometry. 

It is evident  that the code is not generally correctable. However, since superradiance changes the error model, the code becomes correctable. The $T_1$ error changes to superradiance decay; namely, a coherent decay of both the first and the second qubits,  as described by the operator $O=\left(\sigma_{-}^{1}+\sigma_{-}^{2}\right)a^{\dagger}$. 
The third qubit is chosen to be a good qubit. Assuming we have the state $\vert \psi\rangle=a\vert A\rangle+b\vert B\rangle$ before the decay, we get afterward
\[
\begin{split}
\vert \psi_1\rangle=\cos(\theta)[a\vert 111\rangle+b\frac{\vert 010\rangle +\vert 100\rangle }{\sqrt{2}}]\otimes\vert 0_{ph}\rangle \\
+\sin(\theta)[a\frac{\vert 101\rangle+\vert 011\rangle}{\sqrt{2}}+\frac{b}{\sqrt{2}}\frac{2\vert 000\rangle}{\sqrt{2}}]\otimes\vert 1_{ph}\rangle
\end{split}
\]
This error is correctable by measuring $\left(s_{z}^{1}s_{z}^{2}s_{z}^{3}\right)$ and then correcting.


{\it Conclusions and perspectives  ---} 
We have proposed and analyzed the use of EC for increasing the signal to noise ratio of various sensing protocols. 
Due to the very specific characteristics of the sensing signals and the noise model, special EC protocols were designed. 
We have shown that this is a powerful method that could have considerable implications for quantum technologies goals and on precision measurements.

{\it Acknowledgments.--} This work was supported by EU Integrating Project DIADEMS. We thank Roee Ozeri, and Nadav Katz for useful discussions.


\newpage

\begin{widetext}
\section{Supplementary material}

In the following sections, we provide a detailed introduction to the methods used to derive the results presented above, and discuss some of the more complex, less intuitive notions applied in this article. The following topics are presented:

1.1) {\bf The Three qubits Raman transition scheme for EC} 1.2) {\bf EC in the strong noise limit} 1.3) {\bf Spin spin interactions scheme} 1.4) {\bf Simulation of the Sorenson-Molmer scheme} 1.5) {\bf Red side band schemes}\\

2) {\bf Multiple level systems for EC and sensing} 2.1) {\bf A general application scheme} 2.2) {\bf Flip-flops where both spins dissipate}\\

3) {\bf The generality of $T_1$ noise} 3.1) {\bf Proving the generality statements}\\

4) {\bf Methods for NV center schemes}

\subsection{1.1 The three qubits EC mechanism for $T_1$ errors }

Here we provide a detailed explanation of the EC mechanism presented in the 'prolonging $T_1$' section above, for the Raman transition between the two three-qubits states $\ket{A}=\frac{\ket{00}+\ket{11}}{\sqrt{2}}\ket{0}$ and $\ket{C}=\frac{\ket{00}-\ket{11}}{\sqrt{2}}\ket{1}$ through the intermediate state $\ket{B}=\frac{\ket{00}-\ket{11}}{\sqrt{2}}\ket{0}$.
The transition is carried out by lasers shone on the qubits, under the assumption that transition to the $\frac{\ket{00}+\ket{11}}{\sqrt{2}}\ket{1}$ state has been suppressed by the creation of a large energy gap from this state. 

The key parameters are the two Rabi frequencies $g,\Omega$ and the detuning of the two lasers (which are identical) $\delta$. Note that the population of the intermediate state ($\ket{B}$) is proportional to $\epsilon=\frac{\Omega}{\delta} \ll 1$.

Let us assume that, at some time t, we reach the following  state:
\begin{equation}
\left(\beta\vert00+11\rangle\vert0\rangle+\alpha\vert00-11\rangle\vert1\rangle+e^{it\delta}\epsilon\vert00-11\rangle\vert0\rangle\right)\vert n=0\rangle
\end{equation}
A photon is emitted, resulting in:
\begin{eqnarray*}
 & \left(\alpha\vert00\rangle\vert1\rangle+(\beta+e^{it\delta}\epsilon)\vert00\rangle\vert0\rangle\right)\vert n=0\rangle+\\
 & \left(-\alpha\vert01\rangle\vert1\rangle+(\beta-e^{it\delta}\epsilon)\vert01\rangle\vert0\rangle\right)\vert n=1\rangle,
\end{eqnarray*}
where $e^{it\delta}$ is the fast rotating phase of the middle Raman state, and we assume that the emission of the photon is fast relative to the sensing Hamiltonian.

To correct this we measure the spin correlation operator $S_z^1 S_z^2$ : If the first two qubits turn out to be in the same state, we make the correction: $ $$\vert00\rangle\vert0\rangle\rightarrow\vert00+11\rangle\vert0\rangle$,
$\vert00\rangle\vert1\rangle\rightarrow\vert00-11\rangle\vert1\rangle$,  otherwise we make: $\vert01\rangle\vert0\rangle\rightarrow\vert00+11\rangle\vert0\rangle$, $\vert01\rangle\vert1\rangle\rightarrow-\vert00-11\rangle\vert1\rangle$
and thus we get:
\[
 (\beta+e^{it\delta}\epsilon)\vert00+11\rangle\vert0\rangle+\alpha\vert00-11\rangle\vert1\rangle
\]
or
\[
 (\beta-e^{it\delta}\epsilon)\vert00+11\rangle\vert0\rangle+\alpha\vert00-11\rangle\vert1\rangle
\]
Because the times between subsequent decays varies, and since $\langle\delta t\rangle\simeq\delta T_1\gg 1$, the phase $e^{it\delta}$ is essentially random, resulting in a random walk of $\beta$ in 2D, i.e.: $\langle\beta\rangle=0$ and $\langle\vert\beta^2(t)\vert\rangle=\frac{m\epsilon^2}{2}$ where m is the number of measurement circles until time t, and $\epsilon$ is the size of one step. The factor 2 comes from the dimensionality of the random walk. 

From this we get the STD (standard deviation) of $\sigma_{\beta}=\frac{\sqrt{m}\epsilon}{\sqrt{2}}$. Note that the random walk also occurs when no dissipation is measured, so we have $m=\frac{t}{T_1}$. Now we define $T_2^*$, the (new) decoherence time with EC, as the time after which $\beta$ can now longer be separated from the induced noise, that is $\sigma_{\beta}\simeq 1$. This yields
$1=\sqrt{\frac{T_2^*}{T_1}}\epsilon$ or $T_2^*=T_1/\epsilon^2$. As seen above, due to the Raman transition rotation frequency $\omega_{eff}=\frac{g\Omega}{\delta}$, the new accuracy is 
\[
\delta g=\frac{\Omega}{\delta}\delta\omega_{eff}\propto\frac{1}{\epsilon \sqrt{T_2^*}}=\frac{\epsilon}{\epsilon \sqrt{T_1}}= \frac{1}{\sqrt{T_1}},
\]
up to a factor of the order of unity.

\subsection{1.2 Working in the Strong Noise regime}

In a frequency measurement, we can achieve the sensitivity of \cite{martin}:
\begin{equation}
\delta g=\frac{1}{2}\frac{\sqrt{e^{t/T_{1}}-\cos^{2}\left(2gt\right)}}{\sqrt{nTt}\sqrt{\sin^{2}\left(2gt\right)}}
	\label{dg_of_t_sup}
\end{equation}
 as function of the measurement time. To find the optimal time we look for a time such that $\sin^{2}(2gt)\simeq1$. This will give us essentially $\delta g\propto\frac{1}{\sqrt{T_{1} T n}}$ as can be seen in \cite{martin}, where T is the total experiment time and n is the number of, simultaneously running, separate systems, but, in the strong noise regime, $T_{1}\cdot g \ll 1$. In other words, we can only work in the $gt \ll 1$ regime, where $\sin^{2}(2gt)\simeq (2gt)^{2}.$  By substituting in [\ref{dg_of_t_sup}] we get 
\begin{equation}
\delta g=\frac{1}{2}\frac{\sqrt{e^{t/T_{1}}-1+4\left(2gt\right)^{2}}}{\sqrt{nTt}\left(2gt\right)},
	\label{Sdg_of_t_sup}
\end{equation}
differentiating according to t and solving gives us $t_{max}>T_{1}$. Since we cannot prolong the measurement for so long,
we will choose the longest time possible ,i.e. $t=T_{1}$  giving us:
\[
\delta g_{strong}\simeq\frac{1}{2}\frac{\sqrt{e-1}}{\sqrt{nT}2g\left(T_{1}^{3/2}\right)}\propto T_{1}^{-3/2}
\]
This is not good since it increases the error by factor $\frac{1}{gT_1}  \gg 1.$

If we use the three-qubits-scheme for error correction described above in this system, we reduce the frequency to the new $\omega_{eff}=\frac{\Omega g}{\delta}=g\epsilon$, but we can prolong the decoherence, getting $T_2^*  \gg T_1$. Specifically we have $T_2^* \epsilon^2\propto T_1,$ and thus can choose $T_2^* \omega_{eff}=g\epsilon T_2^*=g T_1/\epsilon \gg 1$ and so we go back to the conditions at the beginning of this section, once again achieving the sensitivity described in eq.[\ref{dg_of_t_sup}], but with new parameters:
\[
\delta g=\frac{\delta}{\Omega}\cdot \delta \frac{g\Omega}{\delta}\simeq\frac{\delta}{2\Omega}\frac{\sqrt{e^{t/T_{2}^*}-\cos^{2}\left(2\frac{g\Omega}{\delta}t\right)}}{\sqrt{nTt}\sqrt{\sin^{2}\left(2\frac{g\Omega}{\delta}t\right)}}
\]
where we replaced the old $T_1$ of the system to the $T_2^*$ of the corrected system, and plugged in the Raman transition frequency instead of the simple system's g.

Now, as implied above, we can choose $t$ so that $\sin^{2}\left(2\frac{g\Omega}{\delta}T_{2}\right)= \sin^{2}(2gT_{1}\frac{\delta}{\Omega})=1.$ 
Noting that $\delta g=\left(\frac{df(g,x)}{dg}\right)^{-1}\delta f(g,x) $ (since we assume $\delta x=0$), we get the accuracy
\[
\delta g_{cor}=\frac{\sqrt{e}\delta}{2\Omega}\frac{1}{\sqrt{T_{2} T n}}=\frac{1}{\sqrt{T_{1} T n}},
\]
just as in \cite{martin}. Checking the relative accuracy we get:
\[
 \frac{\delta g_{cor}}{\delta g_{strong}}\propto T_{1}g \ll 1
\]
\subsection {1.3 Sensing spin spin interactions - and flip flop pattern}
We describe in detail error corrections when measuring filp-flop type interaction.
The code we need to correct is:
$\ket{A}=\left( \vert \uparrow 0 0 \rangle + \vert \downarrow 11 \rangle  \right)/\sqrt 2$, $\ket{B}=\left( \vert \downarrow 1 0 \rangle + \vert \uparrow 01 \rangle  \right)/\sqrt 2,$ where the first qubit is subject to $T_1$ noise (i.e. dissipation), and the signal flip-flops between this qubit and a second 'good' qubit, that is: $H_s=\Omega(\sigma_+^1\sigma_-^2+h.c.) $ where $\sigma$ are the Pauli operators.
The third qubit is an auxiliary 'good' qubit.
The Hamiltonian in the code space, is:
\[
H=\Omega(\ket{B}\bra{A}+\ket{A}\bra{B})
\]
which induces rotation between the two code states. 

Suppose we reach the following  state:
\[
\ket{\psi}=a\frac{\vert \uparrow 0 0 \rangle + \vert \downarrow 11 \rangle}{\sqrt2}+b\frac{\vert \downarrow 1 0 \rangle + \vert \uparrow 01 \rangle}{\sqrt2},
\]
and then a dissipation takes place, with amplitude $w$ (where $w \ll 1$), bringing us to the state:
\[
\ket{\psi}=\frac{w}{\sqrt{2}}\left(a\vert \downarrow 0 0 \rangle +b \vert \downarrow 01 \rangle\right)\otimes\ket{1_{ph}} + \left(\frac{ a\vert \downarrow 11 \rangle+\vert \downarrow 1 0 \rangle}{\sqrt2}+\frac{\sqrt(1-w^2)}{\sqrt2}\left[a\vert \uparrow 0 0 \rangle +b \vert \uparrow 01 \rangle\right]\right)\otimes\ket{0_{ph}}.
\]
Now we measure the operator $S_z^1S_z^2$ ($S_z^i$ is the spin, in the z direction, of the i-th qubit), thus casting the code into the state (up to normalization):
\[
\ket{\psi}=\left(a\vert \downarrow 0 0 \rangle +b \vert \downarrow 01 \rangle\right)\otimes\ket{1_{ph}} 
\]
or
\[
\ket{\psi}= \left(\frac{ a\vert \downarrow 11 \rangle+\vert \downarrow 1 0 \rangle}{\sqrt2}+\frac{\sqrt(1-w^2)}{\sqrt2}\left[a\vert \uparrow 0 0 \rangle +b \vert \uparrow 01 \rangle\right]\right)\otimes\ket{0_{ph}}.
\]
In the first case, the error can be corrected by the pulses $ \ket{ \downarrow 0 0}\to \frac{\ket{ \uparrow 0 0}+\ket{ \downarrow 11}}{\sqrt{2}}$ and $\ket{ \downarrow 0 1}\to \frac{\ket{ \uparrow 0 1}+\ket{ \downarrow 10}}{\sqrt{2}}.$
In the second case another measurement is needed; specifically measuring the operator $S_z^2$ will transform the state into either:
\[
\ket{\psi}= \left( a\vert \downarrow 11 \rangle+\vert \downarrow 1 0 \rangle\right)\otimes\ket{0_{ph}}
\]
or 
\[
\ket{\psi}= \left(a\vert \uparrow 0 0 \rangle +b \vert \uparrow 01 \rangle\right)\otimes\ket{0_{ph}},
\]
and in both cases completing the correction should be straight forward.

\subsubsection{1.3.1 Ramsey with flip flops}
Another possibility for EC while sensing flip-flops is to use Ramsey interferometry, for example in the following scheme. For the code we use three qubits. The first qubit state is denoted by ${\ket{u};\ket{d}}$, this qubit is undergoing dissipation. The second qubit, whose state is denoted by $\ket{\uparrow};\ket{\downarrow}$ is a 'good' qubit. Flip-flops occur between the first  and the second qubits. The third qubit is a 'good' auxiliary qubit whose state is denoted by $\ket{0};\ket{1}$.
The Hamiltonian will take the form of $H_s=\Omega(\sigma_-^1\sigma_+^2+h.c.)+\nu(\sigma_z^3)$

The code will be: $ \{\left|A\right\rangle =\frac{\left|u\downarrow\right\rangle +\left|d\uparrow\right\rangle }{\sqrt{2}}\otimes\left|1\right\rangle ,
\{\left|C\right\rangle =\frac{\left|u\downarrow\right\rangle -\left|d\uparrow\right\rangle }{\sqrt{2}}\otimes\left|0\right\rangle\}$ 
and we get the sensing $H_{s}=(\nu+\Omega)\vert A\rangle\langle A\vert$ which allows for Ramsey interferometry for measuring $\Omega$. For the measurement, apply a field $\hat{B}=B_0\sigma_z^3\cdot\cos(\omega\cdot t)$ where $\omega=\Omega+\nu+\delta$ is well-known. Note that this field works only on the third qubit. Apply $\hat{B}$ for a $\pi/2$ pulse on state $\ket{A}$, then wait while making error corrections (every $\tau$ seconds) for time t, then apply $\hat{B}$ for another $\pi/2$ pulse. Measuring the probability of being in state $\ket{C}$ will give the expected cosine signal.

EC procedure: assume we start in the state $\vert \psi\rangle=a\vert A\rangle+b\vert C\rangle$. A dissipation occurs on the first bit, taking us to:
\[
\begin{split}
\vert \psi_1\rangle=\sin(\theta) \frac{a\vert d\downarrow1\rangle+b\vert d\downarrow0\rangle}{\sqrt{2}}\otimes\vert 1_{ph}\rangle +\\ 
[\cos(\theta) \frac{a\vert u\downarrow1\rangle+b\vert u\downarrow0\rangle}{\sqrt{2}}+ \frac{a\vert d\uparrow1\rangle- b\vert d\uparrow0\rangle}{\sqrt{2}}]\otimes\vert 0_{ph}\rangle.
\end{split}
\]
By measuring the (local) operators $S_z^1$ and $S_z^2$ we can separate the system to 
\[
\vert \psi_{-1,-1}\rangle =\frac{a\vert d\downarrow1\rangle+b\vert d\downarrow0\rangle}{\sqrt{2}}
\]
or
\[
\vert \psi_{1,-1}\rangle =\frac{a\vert u\downarrow1\rangle+b\vert u\downarrow0\rangle}{\sqrt{2}} 
\]
or
\[
\vert \psi_{-1,1}\rangle =\frac{a\vert d\uparrow1\rangle- b\vert d\uparrow0\rangle}{\sqrt{2}},
\]
where $\ket{\psi_{-1,1}}$ fits the measurement results of $<S_z^1>=-1/2,<S_z^2>=1/2$, and so forth. From here correcting is straight forward. The only errors left are from the evolution 
$\frac{a\vert d\uparrow1\rangle- b\vert d\uparrow0\rangle}{\sqrt{2}} \leftrightarrow \frac{a\vert u\downarrow1\rangle-b\vert u\downarrow0\rangle}{\sqrt{2}}$ occurring during the dissipation process, which are second order in the noise parameter, giving us $T_2^*\propto \frac{\tau}{T_1}\cdot \frac{1}{T_1}$.

\subsection{1.4 Sorenson Molmer scheme simulation }

\begin{figure}
   \centering
    \includegraphics[scale=0.3]{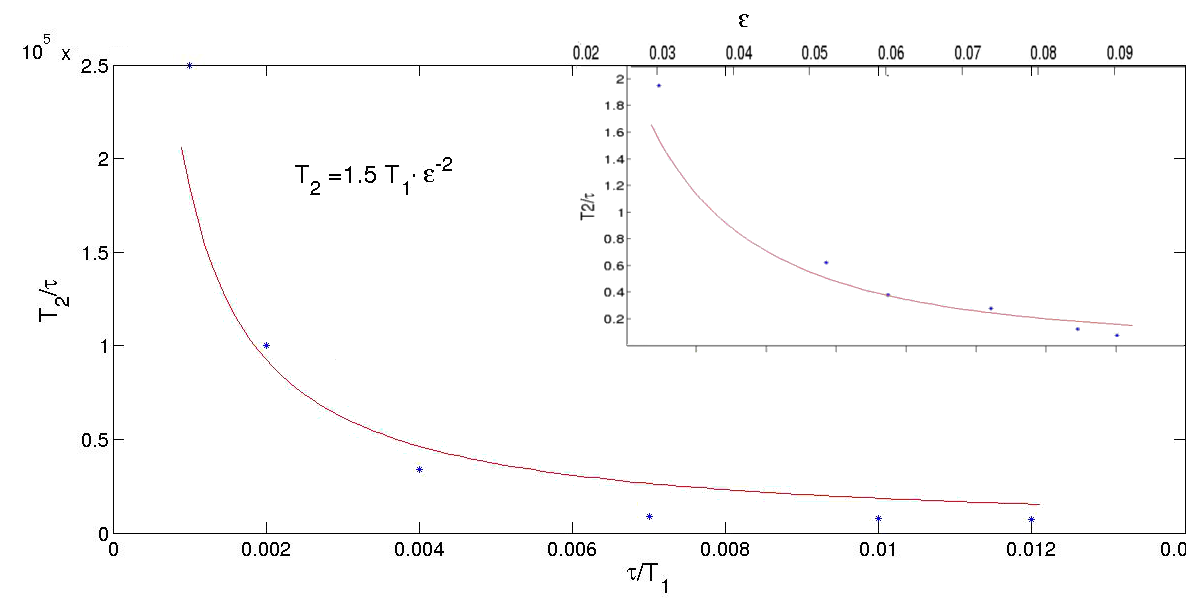}
		\caption {Sorenson Molmer simulation}
	{The graph depicts the dephasing time with error correction ($T_2$) of the Sorenson Molmer simulation system, estimated from the simulation as a function of $\frac{\tau}{T_1}$ and of $\epsilon=\frac{g}{\delta}$ (in the inset). Here $T_1$ is the original decay rate of the system (i.e. without EC). The illustrated curves are described by the formula $T_2=1.5 \frac{T_1}{\epsilon^2}$ a qualitative agreement can be seen.
	Note that although $\tau$ does not appear in the final formula, the formula is only valid in the $\frac{\tau}{T_1}<<1$ limit} 
	\label{MS1}
\end{figure}

The Sorenson Molmer scheme we proposed is not as simple as it might appear, since  it involves a Raman transition through a state which is outside the code. Here, unlike the original scheme, the EC sequence do not induce an error every time (this is why this scheme works better). Nevertheless, every time an emission of a photon is measured, a small error is induced through the correction sequence; in essence this error occurs since we send the population of the intermediate state to 0. 

As the intermediate state is a fast oscillating state, whose phase relative to the other states depends on initial conditions, we assumed that each such correction would effectively send the system to the beginning of the last oscillation cycle of the intermediate state, causing a $1/\delta$ delay (this happens on average every $T_1$, resulting in the very small change of $T_1/\delta$ which can be accounted for exactly). 

Also we assumed that every time we decay, and correct that decay, we effectively lose half the time since last EC sequence, as this is the average time in the lower state in which the sensing Hamiltonian  does not operate. In addition, we note that the eigenvectors of the Hamiltonian are of the form $g\ket{A}-\Omega\ket{C}$ and $g\ket{A}+\Omega\ket{C}+\epsilon\ket{B}$, which are populated in the Raman transition process, and $\epsilon \left(g\ket{A}+\Omega\ket{C}\right)+\ket{B}$. This latter state should only have a population of order $\epsilon$, and  any excessive population means that the population is trapped in the intermediate $\ket{B}$ state and causes degradation of the signal. For this reason we had to assume that the EC does not cause a population transfer to this last state. All these assumptions were validated by simulation. A typical output is visualized in fig. \ref{MS}  in the paper, showing that EC vastly increases the accuracy. 

Furthermore, in fig. \ref{MS1}, the dependency of the system's $T_2^*$ on different parameters is shown, as calculated from the results of the simulation. The fit is rather crude and bears only qualitative resemblance to the expected $T_2^*=1.5\cdot T_1 \epsilon^-2$ curve. This can be attributed to four factors: 1) given our limited resources, each point on the graph shows the average over the results of only a few runs, rather than few thousands, as ideally it should. 2) The actual results depend to a great extent on the number of times in which the emission of a photon was actually detected 3) $T_2^*$ were only estimated up to a factor of order unity rather than calculated exactly 4) higher order contributions were ignored when calculating the expected result.

\section{1.5 \  Side band interaction  (and Molmer-Sorenson)}

Dissipation takes place on the first bit. by setting $\epsilon=\frac{\sqrt{g_1^2+g_2^2}}{\delta}$ we have the typical state of
$\vert \psi\rangle=a\vert A\rangle+c\vert C\rangle+e^{i\phi_B}\epsilon\vert B\rangle$ and, in the fast ECS limit, dissipation bring us to:
\begin{eqnarray}
\textstyle
\vert \psi\rangle =a\frac{(1-\tau\Gamma)\vert \uparrow00\rangle+\vert \downarrow11\rangle}{\sqrt{2}}\vert 0_{vib}\rangle+c\frac{\vert \downarrow10\rangle+(1-\tau\Gamma)\vert \uparrow01\rangle}{\sqrt{2}}\vert 0_{vib}\rangle \nonumber \\
+e^{i\phi_B}\epsilon\frac{\vert \downarrow00\rangle+\vert \downarrow01\rangle}{\sqrt{2}}\vert 1_{vib}\rangle+
\tau\Gamma(\frac{a\vert \downarrow00\rangle+b\vert \downarrow00\rangle}{\sqrt{2}}\vert 0_{vib}\rangle)
\end{eqnarray}
It should be possible to design a measurement that will separate the state, depending on the measurement outcome, in to:
\[
\vert \psi_1\rangle =\frac{a\vert \downarrow11\rangle\vert 0_{vib}\rangle+c\vert \downarrow10\rangle\vert 0_{vib}\rangle+e^{i\phi_B}\epsilon\vert \downarrow00\rangle\vert 1_{vib}\rangle}{\sqrt{2}} 
\]
or 
\[
\vert \psi_2\rangle =\frac{a\vert \uparrow00\rangle\vert 0_{vib}\rangle+c\vert u\uparrow01\rangle\vert 0_{vib}\rangle+e^{i\phi_B}\epsilon\vert \downarrow01\rangle\vert 1_{vib}\rangle}{\sqrt{2}} 
\]
or 
\[
\vert \psi_3\rangle =\frac{a\vert \downarrow00\rangle\vert 0_{vib}\rangle- c\vert \downarrow00\rangle\vert 0_{vib}\rangle}{\sqrt{2}} 
\]

Correcting this will amount to reducing the effective measurement time by $\tau/2$ in the last case, and to errors of order $\Gamma\tau\epsilon^2$ which is third order. In other words, we achieved enlarged $T_2$ and enhanced precision, as was also validated by the simulation presented in the previous section.

\section{2.\ \  Multi-level systems, sensing ,  and EC}
In this paper we mainly assumed that we were dealing with physical qubits; i.e. each distinct qubit is also a distinct atom, spinor or, ion (and so forth). 
The main implication of this assumption is some loss of generality in the possible relation of the sensing Hamiltonian to the error model: for example, when we say (under 'prolonging $T_1$ ') that the most obvious way to connect two states which are more than one step apart is a Raman transition, we are obviously referring to some natural, physical, partition of the qubits.

If we have two states say, $\ket{0000}$ and $\ket{1000}$ connected by some field, we can always denote $\ket{dddd}=\ket{0000}$ and $\ket{uuuu}=\ket{1000}$ and now the same field connects these two seemingly very different states. Of course, such notions are superficial and only cause confusion because they have no real implications, since the errors will now also  connect very different states. In other words, when we measure 'distances' between code states we should really measure them relative to the possible errors. As we noted above, all qubits that are sensitive to the measured field must be also sensitive to errors (otherwise we will use only 'good' qubits and need no EC). This notion is always true, and this is why our 'physicality' assumption is a very good one.

Nevertheless, when describing multilevel systems in terms of 'good' and 'noisy' qubits,  a simple system might still need to be described by very complicated qubits structures. Thus for some systems, the obvious error models might differ from the ones we refer to in this article. ( changing the errors is interchangeable with  changing the sensing Hamiltonian  as regards EC because the orthogonality relations between the two is all that matters).

In practice, finding systems with error models which are correctable is not easy, as the added complexity of the systems tends to cause more evolved errors rather then simply different ones. Still, the possibility exists, and two examples are given below.

\subsection{2.1 Using multiple level systems}
One way to achieve better performance from our EC models is by using systems that decay to states other than the ground states. 
On the down side, such systems will undergo decay from all their subsequent states rather than only 
from the ''up'' state of the qubits. On the up side, each state will decay into a different orthogonal state,
enabling additional freedom in separating and correcting the errors.

We use atomic states in the example.
Denote $\ket{ 1}=\ket{ J=2;m=2}, \ket{ 0}=\ket{ J=2;m=1}$ and
$ \ket{ u}=\ket{ J=1;m=1}, \ket{ d}=\ket{ J=1;m=0}.$
The decay is thus either $\ket{1}\to\ket{u}\otimes\ket{1_{phA}}$ or $\ket{0}\to\ket{d}\otimes\ket{1_{phB}},$ where all the different photons populations contribute to orthogonal states of the environment; i.e. the decays resulting in different photons are not coherent.

We will use a code made up of the two-atom states:
\[
\frac{\ket{ 11}+\ket{ 00}}{\sqrt2},\frac{\ket{ 01}+\ket{ 10}}{\sqrt2}.
\]
The sensing Hamiltonian will be $H_s=g\frac{\ket{ 11}+\ket{ 00}}{\sqrt2}\frac{\bra{ 01}+\bra{ 10}}{\sqrt2}+h.c.$ which can be achieved by 
means of magnetic field in the $\hat{x}$ direction, acting on each atom separately.
Note that the decay on different atoms is non-coherent in other words there are different phase-shifts for different atoms as well as for different decays.

Starting with:
\[
\ket{ \psi}=a\frac{\ket{ 11}+\ket{ 00}}{\sqrt2} +b\frac{\ket{ 01}+\ket{ 10}}{\sqrt2} ,
\]
A decay brings us to
\[
\begin{split}
\ket{ \psi'}=w\left(a\frac{\ket{ 11}+\ket{ 00}}{\sqrt2} +b\frac{\ket{ 01}+\ket{ 10}}{\sqrt2}\right) \\
 + A\frac{a\ket{ u1}+b\ket{ u0}}{\sqrt2}\otimes\ket{1_{phA}}+B\frac{b\ket{ d1}+a\ket{ d0}}{\sqrt2}\otimes\ket{1_{phB}}\\
 + C\frac{a\ket{ 1u}+b\ket{ 0u}}{\sqrt2}\otimes\ket{1_{phC}}+D\frac{a\ket{ 0d}+\ket{ 1d}}{\sqrt2}\otimes\ket{1_{phD}}
\end{split}
\]
where $A,B,C,D$ are the amplitudes to emit the 4 respective possible distinct photons.
Measuring each atom's J value will either correct the errors or bring us (w.l.o.g.), up to normalization, to:

\[
A\frac{a\ket{ u1}+b\ket{ u0}}{\sqrt2}\otimes\ket{1_{phA}}+B\frac{b\ket{ d1}+a\ket{ d0}}{\sqrt2}\otimes\ket{1_{phB}}.
\]
Now measuring the $J_z$ state of the first atom (that is, its m value); i.e. measuring in the $\left\{\ket{u} ; \ket{d}\right\}$ basis, we get (w.l.o.g.):

\[
\left(a\ket{ u1}+b\ket{ u0}\right)\otimes\ket{1_{phA}}
\]
where now the photon number has no significance, and can be ignored. Now make $\ket{ u1} \to \frac{\ket{ 11}+\ket{ 00}}{\sqrt2} $ and $\ket{ u0} \to \frac{\ket{ 10}+\ket{ 01}}{\sqrt2} $ and the error is fully corrected. Note that this method, in fact, enables the measurement of generic magnetic field.

\subsection{2.2 Measurement of flip-flops where both atoms decay} 
This can be done by using multiple level systems.  We use two prob atoms in the J=1 and J=0 states, and one good qubit in some other states $\left\{\ket{u} ; \ket{d}\right\}$. 
Denote $\ket{ 1}=\ket{ J=1;m=1}, \ket{ 0}=\ket{ J=1;m=0}$ and
$ \ket{ !}=\ket{ J=0;m=0}$. We  assume flip flops of the two qubits (i.e. $H_s=g\ket{10}\bra{01} +h.c.$),
both of which may undergo decay in the form:
\[
	\ket{ 1}\to \ket{!}\otimes\ket{1_{phA}}\ \   ;\ \  \ket{ 0}\to \ket{ !}\otimes\ket{1_{phB}},
\]
where the two modes of decay are non coherent, and the decays of different atoms are also non-coherent.
We use the three-atom code:

 \[
\frac{\ket{ 10u}+\ket{ 01d}}{\sqrt2},\frac{\ket{ 01u}+\ket{ 10d}}{\sqrt2}.
\]
Starting with :
\[
\ket{ \psi}=a\frac{\ket{ 10u}+\ket{ 01d}}{\sqrt2}+b\frac{\ket{ 01u}+\ket{ 10d}}{\sqrt2},
\]
a decay brings us to
\[
\begin{split}
\ket{ \psi'}=w\left(a\frac{\ket{ 10u}+\ket{ 01d}}{\sqrt2}+b\frac{\ket{ 01u}+\ket{ 10d}}{\sqrt2}\right) \\
 + A\left(a\frac{\ket{ !0u}+\ket{ !0d}}{\sqrt2}\otimes\ket{1_{phA}}+B\frac{\ket{ !1u}+\ket{ !1d}}{\sqrt2}\otimes\ket{1_{phB}}\right)\\
 + C\left(a\frac{\ket{ 0!u}+\ket{ 0!d}}{\sqrt2}\otimes\ket{1_{phC}}+D\frac{\ket{ 1!u}+\ket{ 1!d}}{\sqrt2}\otimes\ket{1_{phD}}\right)
\end{split}
\]
where $A,B,C,D$ are the amplitudes to emit the 4 respective possible distinct photons.
Measuring the J state of each atoms, we either correct the error or reach (w.l.o.g.) as the state:
\[
\left(A\frac{\ket{ !0u}+\ket{ !0d}}{\sqrt2}\otimes\ket{1_{phA}}+B\frac{\ket{ !1u}+\ket{ !1d}}{\sqrt2}\otimes\ket{1_{phB}}\right),
\]
now, measuring the M state of the second atom brings us to:
\[
\left(a\ket{ !0u}+b\ket{ !0d}\right)
\]
or
\[
\left(a\ket{ !1d}+b\ket{ !1u}\right)
\]
where we have left out the photons, as they are now unimportant. Correcting from here is straight forward.

\section{3.\ \  $T_1$ decay error  versus General Error}
In the language of error correction it is customary to refer to two kinds of errors; phase-flip (i.e. errors proportional to the $\sigma_z$ Pauli operator) and bit-flip (i.e. $\sigma_x$ errors). This is because these two errors are simple to comprehend, and span all possible errors, in the sense that if  a bit-flip and  a phase-flip can be corrected any general error can be corrected. These kind of codes are 'fully correctable' and can be corrected for 'general errors'.

Specifically this means that in the former case we can also correct errors induced by decay. Sometimes the converse is also held to be true, but it is not. Below we demonstrate, by means of an example, that correcting decay errors does not require the power to correct any error. We also present and explain why one might, naively, believe the converse.

For clarity, we define errors induced by decay (also refereed to as $T_1$ errors). These errors are caused by coupling the system to the environment via the term $a_{i}^\dagger\sigma_- + h.c,$ where $a_i$ is the annihilation operator of the i-th mode of a of the environment. This error, when one traces-out the state of the cavity (i.e. the environment), causes decoherence. The rate of the error is defined by the typical time scale, denoted $T_1$, after which a decay is likely to occur. In what follows we will suppose, for simplicity, that the photon mode was originally found in the $\ket{n=0}$ state.

\subsection {3.1 $T_1$ errors are not general errors}
To prove this consider the following eight qubit system, which can be corrected  for bit-flip and $T_1$ errors on each qubit, but not for the effects of phase-flip errors. Denote $\ket{+}=\frac{\ket{1010}+\ket{0101}}{\sqrt{2}} $ and $\ket{-}=\frac{\ket{1010}-\ket{0101}}{\sqrt{2}},$
we shell define our code states, and our initial state, to be:
\[
\begin {split}
\ket{A}=\ket{+}\ket{+} ; \ket{B}=\ket{-}\ket{-}\\
\ket{\psi}=a\ket{A}+b\ket{B}
\end{split}
\]

{\bf Bit -flip error}, occurring with amplitude $\epsilon \ll 1$ bring us to,
\[
\begin{split}
\ket{\psi'}=(1-4\epsilon^2)\ket{\psi}+
\epsilon\left(a\frac{\ket{0010}+\ket{1101}}{\sqrt{2}}\ket{+}+b\frac{\ket{0010}-\ket{1101}}{\sqrt{2}}\ket{-}\right)\\
+...+
\epsilon\left(a\ket{+}\frac{\ket{1011}+\ket{0100}}{\sqrt{2}}+b\ket{-}\frac{\ket{1011}-\ket{0100}}{\sqrt{2}}\right)
+higher\ order
\end{split}
\]
Note that while we assume all qubits might have errors, we also assume that only one qubit eror at the same time; that is, we assume that the probability of measuring an error is proportional to some small parameter (and thus multiple simultaneous errors are second order). 

It is evident that this error can be corrected by measuring the adjacent-qubits correlation operators $S_z^{i}S_z^{i+1}$. Two neighboring qubits with the same sign tells us that an error has occurred, as well as revealing its location. 

{\bf Phase-flip errors}, however, cannot be corrected, since a phase-flip on the the first four qubits coincides with the phase-flip on the last four, in a nasty way:
\[
\begin{split}
\ket{\psi'}=(1-4\epsilon^2)\ket{\psi}+
2\epsilon\left(a\ket{-}\ket{+}+b\ket{+}\ket{-}\right)\\
+2\epsilon\left(a\ket{+}\ket{-}+b\ket{-}\ket{+}\right)
\end{split}
\]
which is
\[
\ket{\psi'}=(1-4\epsilon^2)\ket{\psi}+
2\epsilon\left((a+b)\ket{-}\ket{+}+(b+a)\ket{+}\ket{-}\right)
\]
and since the separation between the populations, that is a and b, was destroyed, the error cannot be corrected.

{\bf Decay errors}  occurring on the first bit  with amplitude $\epsilon$, will bring us to the  following state: 
\[
\begin{split}
\ket{\psi'}=\left(\frac{(1-\epsilon^2/\sqrt{2}) a \ket{1010}+a \ket{0101}}{\sqrt{2}}\ket{+}+\frac{(1-2\epsilon^2/\sqrt{2})b \ket{1010}- b \ket{0101}}{\sqrt{2}}\ket{-}\right)\ket{n=0}\\
+\epsilon\left(\frac{a}{\sqrt{2}}\ket{0010}\ket{+}+\frac{b}{\sqrt{2}}\ket{0010}\ket{-}\right)\ket{n_1=1}
\end{split}
\]
now, if we allow for decay on any of the bits we get
\[
\begin{split}
\ket{\psi'}=(1-2\epsilon^2)\ket{\psi}\ket{n=0}+
\epsilon\left(a\frac{\ket{0010}}{\sqrt{2}}\ket{+}+b\frac{\ket{0010}}{\sqrt{2}}\ket{-}\right)\ket{n_1=1}\\
+...+
\epsilon\left(a\ket{+}\frac{\ket{0100}}{\sqrt{2}}-b\ket{-}\frac{\ket{0100}}{\sqrt{2}}\right)\ket{n_8=1}.
\end{split}
\]
This error can be resolved, again, by measuring the two spins correlation functions. If, for example, we measure the second qubit to be the same as its neighbors, we necessarily arrived at the state
\[
\ket{\psi''}=\left(a\frac{\ket{0001}}{\sqrt{2}}\ket{+}+b\frac{\ket{0001}}{\sqrt{2}}\ket{-}\right)\ket{n_2=1}
\]
which indeed enables us to correct the error. The correction from here is straight forward.
Now, let's assume that we measured no decay, that is, all the adjacent spin correlations turn out negative, in which case we reach the state:
\[
\ket{\psi''}=\ket{\psi}\ket{n=0}
\]
and thus the error has already been corrected. The conclusion is that since in this system it is possible to correct $T_1$ errors, but not phase-flip errors, then correcting $T_1$ errors is not equivalent to correcting any error (that is, to correcting ''general errors'') . 

\subsection {3.2 $T_1$ errors are similar to general errors}
Despite the above demonstration, in many systems $T_1$ errors appear to be almost as bad as general errors; in other words, they are almost as hard to correct.  Evidently, correction of $T_1$ errors implies the correction of bit-flip errors. In addition, since the error is a non Hermitian $\sigma_-$ error, it operates differently on the up and down states. Thus, in order to correct it, the code states need to have the same probability for being in the up state of each qubit. That is, the code states cannot differ in the probability of finding oneself in the up state of any of the qubits.
Suppose for example the following state:
\[
\ket{\psi}=a\ket{111}+b\ket{000}
\]
Then a decay,  occurring on the first bit for example, will take us to
\[
\ket{\psi}=((1-\epsilon^2/\sqrt{2})a\ket{111}+b\ket{000})\ket{n=0}+\epsilon(a\ket{011})\ket{n=1}.
\]
Evidently, this cannot be corrected.
Also take a look at the code states $\ket{1111}$ and $\frac{\ket{1010}+\ket{0101}}{\sqrt{2}}$. This also is not correctable for the same reason, since the decay rate of each state is different, and thus each correction induces a $\sqrt{2}$ relative factor.

The next obvious system one might check is the following Schor's-like code: denote $\ket{\pm}=\ket{111}\pm\ket{000}$; then the states are $\ket{+}\ket{+}$ and $\ket{-}\ket{-}.$ Assume a decay, for brevity of the first qubit only, and we get from $\ket{\psi}=a\ket{+}\ket{+}+b\ket{-}\ket{-}$ the state
\[
\begin{split}
\ket{\psi}=(a\frac{\ket{000}}{\sqrt{2}}\ket{+}-b\frac{\ket{000}}{\sqrt{2}}\ket{-})\ket{n=0}
+(1-\epsilon^2/\sqrt{2})(a\frac{\ket{111}}{\sqrt{2}}\ket{+}+b\frac{\ket{111}}{\sqrt{2}}\ket{-})\ket{n=0}\\
+\epsilon(a\frac{\ket{011}}{\sqrt{2}}\ket{+}+b\frac{\ket{011}}{\sqrt{2}}\ket{-})\ket{n=1}
\end{split}
\]
By measuring the correlation operators it is easy to verify that a decay here is correctable. But look what happens when no decay was measured:  one can write $\ket{000}=\frac{\ket{+}-\ket{-}}{\sqrt{2}}$ and thus we arrive at the state:
\[
\ket{\psi}=a((1-\epsilon^2/2)\ket{+}+\epsilon^2/2\ket{-})\ket{+}-b(1-\epsilon^2/2)\ket{-}+\epsilon^2/2\ket{-})\ket{-}
\]
This is a phase-flip and we know that phase-flips are not correctable in this system.
It should be pointed out that this is only an ''effective phase-flip'' due to the structure of the system, since in our previous system no such phase-flip occurred.
Also note that in this system,  apparently we need to be able to correct bit-flips, and subsequent bit and phase flips (on the same qubit) - when we measure decay. Furthermore, we have to be able to correct phase-flip when no decay was measured. Altogether this signifies a ''general error''. Misleading us to believe that $T_1$ errors are equivalent to general errors. Note that the ''effective phase-flip'' appears only in second order in $\epsilon$ which is advantageous, since this is also the order of the non correctable ''two bit-flips'' errors. 

\appendix

\setcounter{equation}{0}

\vspace{1ex}

\section {4 Error Correction Protocol for Improving DD} 
In this section we present the error correction protocol applicable for both the models described in the article under Classical Drive Noise and Quantum Noise. In both models, we presented the mapping of the physical space into code states, in Eqs.(~\ref{code_classical}) and the error acts as a bit-flip that maps the states from the code space onto the error space. Denoting by $\vert 0_c,0\rangle$ and $\vert 1_c,1\rangle$ the code states, the error operation will take these states to the states  we  denote by $\vert 1_c,0\rangle$ and $\vert 0_c,1\rangle$, respectively.

Preparing the system in an initial state $\vert\psi(0)\rangle = (\vert 0_c,0\rangle + \vert 1_c,1\rangle)/\sqrt{2}$, the sensing part of the Hamiltonian will take it, in time $t$, to be
\begin{equation}
	\vert \psi(t)\rangle = 
			\frac{\cos\left(\frac{g+\nu}{2}t\right)}{\sqrt{2}}
			(\vert 0_c,0\rangle + \vert 1_c,1\rangle)
			+\frac{i\sin\left(\frac{g+\nu}{2}t\right)}{\sqrt{2}}
			(\vert 0_c,0\rangle - \vert 1_c,1\rangle),
\end{equation}
and $g$ can be deduced by measuring the probability of the system will be still at the initial state, at time $t.$ An error operation will take the system into the error states defined above. We now measure the Hermitian operator
\begin{eqnarray}
	\Sigma_z &=& \vert 0_c,0\rangle\langle 0_c,0\vert  +
				\vert 1_c,1\rangle\langle 1_c,1\vert  - \nonumber \\ &&
				\vert 0_c,1\rangle\langle 0_c,1\vert  -
				\vert 1_c,0\rangle\langle 1_c,0\vert ,
\end{eqnarray}
which returns the eigenvalue $(+1)$ if the system had no error and $(-1)$ if a single error has occurred, without changing the state of system in either case (up to a global irrelevant phase). Having discovered that an error occurred, the state can be corrected by applying the a bit-flip operation. Thus we have outlined a complete error correction protocol for both these models.

\end{widetext}

\end{document}